\documentclass{article}
\usepackage{graphicx}
\usepackage{bm}
\usepackage{color}
\usepackage{amsmath}
\usepackage{amsfonts}
\usepackage{algorithm}
\usepackage{algorithmicx}
\usepackage{algpseudocode}
\usepackage{array}
\usepackage{afterpage,array,rotating}
\usepackage{multirow}
\newcolumntype{Y}{>{\centering\arraybackslash}p{5.5em}}

\newcommand{\nn}{\nonumber}
\usepackage{authblk}
 
\begin{document}

\title{Solving Inequality-Constrained Binary Optimization Problems on Quantum Annealer}

\author[1,2,3]{Kouki Yonaga}
\author[1,2]{Masamichi J. Miyama}
\author[1,2,4]{Masayuki Ohzeki}

\affil[1]{Graduate School of Information Sciences, Tohoku University, Sendai 980-8579, Japan}
\affil[2]{Sigma-i Co., Ltd., Tokyo 108-0075, Japan}
\affil[3]{MathAM-OIL, AIST, Sendai 980-8577, Japan}
\affil[4]{Institute of Innovative Research, Tokyo Institute of Technology, Yokohama 226-8503}
\date{}

\maketitle

\begin{abstract}
    We propose a new method for solving binary optimization problems under inequality constraints using a quantum annealer.
    To deal with inequality constraints, we often use slack variables, as in previous approaches.
    When we use slack variables, we usually conduct a binary expansion, which requires numerous physical qubits.
    Therefore, the problem of the current quantum annealer is limited to a small scale.
    In this study, we employ the alternating direction method of multipliers.
    This approach allows us to deal with various types using constraints in the current quantum annealer without slack variables.
    To test the performance of our algorithm, we use quadratic knapsack problems (QKPs).
    We compared the accuracy obtained by our method with a simulated annealer and the optimization and sampling mode of a D-Wave machine.
    As a result of our experiments, we found that the sampling mode shows the best accuracy.
    We also found that the computational time of our method is faster than that of the exact solver when we tackle various QKPs defined on dense graphs.
    
\end{abstract}
    
\section{\label{sec:introduction}Introduction}    
    Combinatorial optimization problems are essential challenges that emerge in numerous domains such as portfolio optimization \cite{Portfolio}, traffic flow \cite{Traffic}, job-shop scheduling \cite{JobShop}, nurse scheduling \cite{Nurse}, automated guided vehicles \cite{AGV}, and machine learning \cite{ML1}.
    Many researchers have been developing new algorithms to solve these large-sized problems.
    Quantum annealing (QA) is a recently developed technology for solving combinatorial optimization problems \cite{QA_Kadowaki}.
    This technology was initially proposed in academia, inspired by simulated annealing (SA) \cite{SA_Kirkpatrick}.
    With the recent realization of quantum annealers \cite{DW_Berkley, DW_Harris, DW_Johnson}, i.e., D-Wave machines, many researchers have been studying QA for application in industry.
    Thus, QA is attracting significant attention from numerous people in academia and business.
    The current D-Wave machine, D-Wave 2000Q, can minimize the following quadratic cost function:
    \begin{equation}
        \label{eq:cost} 
        E(\bm{x}) = \bm{x}^T Q \bm{x}
    \end{equation}
    where $\bm{x} = \{ 0, 1\}^N$ is an $N$-dimensional vector of binary variables, and $Q$ is an integer or real matrix.
    Eq.(\ref{eq:cost}) and $Q$ are called the quadratic unconstrained binary optimization (QUBO) problem and QUBO matrix, respectively.
    To use the D-Wave 2000Q in practical situations, we represent our task using a QUBO formulation.
    In this study, we assume that our task is given with the following linear constraints:
    \begin{align}
        \label{eq:QP}
        & \underset{{\bm x}} {\text{minimize}} && f(\bm{x}) \nn \\, 
        &\text{subject to} && \bm{F}_l \bm{x} = C_{l} && (l=1,\cdots,L) \nn \\, 
        & && \bm{G}_{m} \bm{x} \leq D_{m} && (m=1,\cdots,M), 
    \end{align}
    where $\bm{F}_l, \bm{G}_m \in \mathbb{Z}^N$, $C_l, D_m \in \mathbb{Z}$, and $f(\bm{x})$ is an objective function given as the QUBO formulation.
    Here, we represent the equality constraints with the penalty terms as follows \cite{QUBO_form}:
    \begin{equation}
        \label{eq:eq_penalty}
        \bm{F}_l \bm{x} = C_l \ \ (l = 1, 2, \cdots L) \ \Leftrightarrow \ \sum_{l=1}^{L} \left( \bm{F}_l \bm{x} - C_l \right)^2, 
    \end{equation}
    Then, adding Eq.(\ref{eq:eq_penalty}) into $f(\bm{x})$, we obtain the QUBO-type cost function.
    In a similar way, the inequality constraints can be written as follows \cite{Slack1, Slack2, Slack3}:
    \begin{equation}
        \label{eq:ineq_penalty}
        \bm{G}_m \bm{x} \le D_m \ \ (m = 1, 2, \cdots M) \ \Leftrightarrow \ \sum_{m=1}^{M} \left( \bm{G}_m \bm{x} - D_m + s_m \right)^2, 
    \end{equation}
    where $s_{m}$ is called the slack variable.
    Thus, the inequality constraints can be represented using the QUBO formulation by the binary expansion $s_m = 1x_1 + 2x_2 + 4x_3 + \cdots $.

    Unfortunately, we can solve only small-sized problems if we apply the slack variables because the binary expansion requires many physical qubits, and D-Wave 2000Q has only approximately 2000 qubits.
    In addition to the slack variable, the embedding techniques limit the problem size that can be solved.
    The physical qubits in the D-Wave 2000Q connect to other qubits in the chimera graph.
    The connection of the hardware, chimera graph, is sparse and differs from that of a logical variable representing the optimization problems.
    Therefore, we use the embedding technique to represent the logical variables on the chimera graph \cite{Ocean_Doc, Embedding_Cai, Embedding_Klymko, Embedding_Okada}.
    The embedding allows us to solve various QUBO problems but uses numerous additional physical bits.
    We can compute only 64 logical variables when the problem is defined on a fully connected graph.
    As a result, the number of logical variables we can use dramatically decreases.
    Thus, it is difficult for the D-Wave 2000Q to deal with inequality constraints.
    
    In this study, we report a new method for solving inequality-constrained binary optimization problems in the D-Wave 2000Q.
    Our algorithm is based on the augmented Lagrangian method and the alternating direction method of multipliers (ADMM) \cite{Aug_Lag1, Aug_Lag2, ADMM1, ADMM2}.
    These approaches allow us to solve the inequality constrained problems without the slack variables.
    Our algorithm applies not only to the D-Wave machine but also to other QUBO solvers.
    The current digital QUBO solvers can deal with more logical variables than the D-Wave 2000Q.
    Therefore, with our method, we can solve larger-sized problems involving the inequality constraints.

    The remainder of this paper is as follows. 
    In Sec.\ref{sec:methods}, we provide an overview of the QA and D-Wave machine.
    Furthermore, we show the augmented Lagrangian method and the main algorithm based on the ADMM.
    In Sec.\ref{sec:experiments}, we describe the test results of our method on quadratic knapsack problems.
    In Sec.\ref{sec:discussion}, we compare the accuracy and computation time obtained by our method and exact optimizers.
    We then discuss the potential superiority of our method over the exact optimizers.
    Finally, we summarize our study in Sec.\ref{sec:summary}.

\section{\label{sec:methods}Methods}
    \subsection{Overview of Quantum Annealing and D-Wave Machine}
        In QA, we set the system, which consists of the target and driving Hamiltonian \cite{QA_Kadowaki, QA_Albash, DW_Doc}.
        The target Hamiltonian includes the Pauli matrices, whose z-components are given as Ising variables as $+1$ and $-1$.
        The target Hamiltonian corresponds to the cost function $E(\bm{x})$ because the Ising variable $s_i$ can be written as $s_i = 2x_i - 1$.
        The driving Hamiltonian introduces quantum fluctuations to the system.
        In the early step of QA, the driving Hamiltonian creates a superposition of all solutions.
        By gradually reducing the influence of the driving Hamiltonian, we obtain the lowest-cost solution for the target Hamiltonian.
        Thus, QA achieves the optimal solution if the annealing time is sufficiently long.
        However, we typically set the annealing time to 20 $\mu$s when actually using the D-Wave 2000Q.
        In addition, it is difficult to remove the effects of noise in the actual system.
        Therefore, the D-Wave 2000Q is used as a sampler, which provides stochastically approximated solutions \cite{Sampler_Benedetti}.
        
        Herein, we introduce postprocessing modes used in the D-Wave 2000Q, i.e., optimization and sampling modes \cite{DW_Doc}.
        The optimization model conducts local updates to the samples obtained through QA.
        Thus, we obtain a set of samples with a lower cost function.
        In sampling mode, the samples obtain using QA are modified into a target Boltzmann distribution, which is defined as 
        \begin{equation}
            P(\bm{x}) = \frac{1}{Z} {\rm exp}\left[ -\beta E(\bm{x}) \right]. 
        \end{equation}
        where $\beta$ is inverse temperature.
        When we take $\beta \rightarrow \infty$, only the lowest-energy samples are obtained.
        By contrast, when $\beta$ moves toward zero, diverse samples are generated from $P(\bm{x})$.

        As mentioned in the previous section, we use the embedding technique.
        Moreover, the unembedding technique is also essential \cite{Ocean_Doc}.
        We use the unembedding to obtain samples on the logical variables after applying QA.
        The D-Wave 2000Q has several unembedding methods, and the default setting employs the majority-vote method.
        In this study, we use the minimize-energy method.
        This method leads us to lower-cost samples by minimizing the local cost function.

    \subsection{Augmented Lagrangian Method}
        We define the cost function including the inequality constraints.
        For simplicity, we consider only the inequality constraints in Eq.(\ref{eq:QP}).
        The inequality constraints can be written using the penalty terms as follows: 
        \begin{equation}
            \label{eq:ineq_cost}
            E_{\rm ineq}(\bm{x}) = f(\bm{x}) + \gamma \sum_{m=1}^M \Theta(\bm{G}_{m}\bm{x}-D_{m}),
        \end{equation}
where $\gamma$ is relatively larger than the objective function $f(\bm{x})$.
        Here, $\Theta(x)$ is the Heaviside step function, which is defined as follows:
        \[
            \Theta(x) = \begin{cases}
            1 & (x >  0) \\
            0 & (x \le 0).
            \end{cases}
        \]
        When $\Theta(\bm{G}_{m}\bm{x}^*-D_{m})$ is zero for ${}^\forall m$, $\bm{x}^*$ is the feasible solution.
        However, the D-Wave 2000Q cannot directly deal with Eq. (\ref{eq:ineq_cost}) because of the Heaviside step function.
        We introduce the augmented Lagrangian method to transform $E_{\rm ineq}(\bm{x})$ into the QUBO formulation \cite{Aug_Lag1, Aug_Lag2}.
        Eq.(\ref{eq:ineq_cost}) can be rewritten as follows:  
        \begin{align}
            & \underset{{\bm x}} {\text{minimize}} && f(\bm{x}) +  \gamma \sum_{m=1}^M \Theta(z_{m}) \nn\\, 
            &\text{subject to} && \bm{G}_{m}\bm{x}-D_{m} = z_{m} && (m=1,\cdots,M), 
        \end{align}
where $\{ z_{m} \} \in \mathbb{Z}^M $ are auxiliary variables.
        We obtain the new cost function $E_{\rm aug}$ with the Lagrangian multipliers and the penalty terms as follows:
        \begin{align}
            \label{eq:aug_cost}
            E_{\rm aug}(\bm{x}, \bm{z}, \bm{\lambda}) &= f(\bm{x}) + \gamma \sum_{m=1}^M \Theta(z_m) \nn \\
             & + \sum_{m=1}^M \lambda_{m} (\bm{G}_{m}\bm{x}-D_{m}-z_{m}) + \frac{\rho}{2}\sum_{m=1}^M \left( \bm{G}_{m}\bm{x}-D_{m} - z_{m}\right)^2, 
        \end{align}
where $\{ \lambda_m \}$ and $\rho$ are the multipliers and coefficients for the penalty terms, respectively.

        \subsection{Main Algorithm}
        To solve $E_{\rm aug}(\bm{x}$, $\bm{z}$, and $\bm{\lambda})$, ADMM is widely used \cite{ADMM1, ADMM2}.
        In ADMM, we update $\bm{x}$, $\bm{z}$, and the multipliers $\bm{\lambda}$ by applying the sequential optimizations as follows:
        \begin{subequations}
        \begin{align}
            &\bm{x}^*[t+1] = \underset{\bm{x}}{\rm argmin}\ E_{\rm aug}(\bm{x}, \bm{z}^*[t+1], \bm{\lambda}[t]), \label{eq:xupdate}\\
            &\bm{z}^*[t+1] = \underset{\bm{z}}{\rm argmin}\ E_{\rm aug}(\bm{x}^*[t+1], \bm{z}, \bm{\lambda}[t]]), \label{eq:zupdate}\\
            &\lambda[t+1] = \lambda[t] + \rho\left( \bm{G}_m^T \bm{x}^*[t+1] - D_m - z^*_{m}[t+1] \right) \ \ (m=1, \cdots, M), \label{eq:lamupdate}
        \end{align}
        \end{subequations}
where $t$ corresponds to the number of iterations.
        By repeating Eqs.(\ref{eq:xupdate})–(\ref{eq:lamupdate}) until convergence, we eventually obtain the optimal solution.
        In this study, we developed a hybrid algorithm that combines ADMM and QA.
        The main difference between the usual ADMM and our hybrid algorithm is the use of a quantum annealer for solving Eq.(\ref{eq:xupdate}).
        After applying QA for $E_{\rm aug}(\bm{x}$, $\bm{z}$, and $\bm{\lambda})$, we obtain the samples $\{ \bm{x}_{\nu} \}$, where $\nu$ is the index for each sample.
        We define the lowest-cost solution $\bm{x}^*_{\rm cost}$ that minimizes $E_{\rm aug}(\bm{x}, \bm{z}, \bm{\lambda})$ as 
        \begin{equation}
            \bm{x}^*_{\rm cost} = \underset{\{ \bm{x}_{\nu} \}}{\rm argmin}\ E_{\rm aug}(\bm{x}, \bm{z}, \bm{\lambda}).
        \end{equation}
        Note that $\bm{x}^*_{\rm cost}$ is not necessarily a feasible solution, but other samples in $\{ \bm{x}_{\nu} \}$ can be feasible.
        Here, we define a feasible solution $\bm{x}^*_{\rm feas}$ that minimizes $f(\bm{x})$ and satisfies the inequality constraints as follows: 
        \begin{equation}
            \bm{x}^*_{\rm feas} = \underset{\{ \bm{x}_{\nu} \}}{\rm argmin}\ f(\bm{x})\ \ \text{s.t.}\ \ \bm{G}_m \bm{x} \le D_m \ \ (m=1,\cdots,M).
        \end{equation}
        We use $\bm{x}^*_{\rm cost}$ to update $\bm{z}$ and $\bm{\lambda}$, whereas $\bm{x}^*_{\rm feas}$ is utilized for searching the feasible solution.
        We show the details of our ADMM algorithm as follows:
        \begin{enumerate}
            \item Initialize the parameters as $\{ z_m \}=0 ,\ \{\lambda_m \} =0$, and $t=1$. \label{alg_initialize}
            \item Apply the embedding for a fully connected graph with size $N$.
            \item Compute the QUBO matrix using Eq.(\ref{eq:aug_cost}): \label{alg_qubo}
            \item Obtain the samples $\{ \bm{x}_{\nu} \}$ by annealing the QUBO matrix. \label{alg_sampling}
            \item Compute $\bm{x}^*_{\rm cost}$ and $\bm{x}^*_{\rm feas}$ using the samples $\{ \bm{x}_{\nu} \}$.
            \item Update $\bm{z}^*$ as $ z_m^* = {\rm min}\left(0,  \bm{G}_m \bm{x}_{\rm cost}^* - D_m \right)\ \ (m=1,\cdots,M)$ \label{alg_zupdate}
            \item Update $\bm{\lambda}$ as $\lambda_m = \lambda_m + \rho\left( \bm{G}_m \bm{x}^*_{\rm cost} - D_m - z^*_m \right)\ \ (m=1,\cdots,M)$ \label{alg_lamupdate}
            \item Check the convergence: When one of the following criteria is satisfied, the calculation is completed. \label{alg_check}
            \begin{enumerate}
                \item $t > t_{\rm max}$
                \item $E_{\rm ineq}(\bm{x}^*_{\rm feas})$ is not improved in $t_{\rm conv}$ steps
                \item $\sqrt{\sum_{m} \left(\bm{G}_m\bm{x}^*_{\rm feas} - D_m - z_m\right)^2 } < \epsilon$
            \end{enumerate} 
            \item[] where $t_{\rm max}$, $t_{\rm conv}$, and $\epsilon$ are predetermined parameters.
            \item $t \leftarrow t + 1$. \label{alg_next}
            \item Iterate (\ref{alg_qubo})-(\ref{alg_next}) until convergence. \label{alg_repeat}
        \end{enumerate}
        Here, the unembedding process is involved in step \ref{alg_sampling}.
        Thus, the unembedding is applied after every sampling, whereas the embedding is conducted once before the iterating part.

        Note the essential points of our ADMM algorithm in the following.
        First, we use the auxiliary variable $\bm{z}$ instead of the slack variable $\bm{s}$.
        This leads to an efficient utilization of D-Wave 2000Q because the binary expansion is not necessary.
        Second, we search for the optimal solution using the sampler.
        Because D-Wave 2000Q is a stochastic sampler, $\bm{x}^*_{\rm feas}$ does not necessarily correspond to the optimal solution even when the ADMM is finished.
        Therefore, we generate many samples using the D-Wave 2000Q and search for a more accurate and feasible solution.

\section{\label{sec:experiments}Experiments}

    \begin{figure}[t]
        \begin{center}
            \includegraphics[clip,width=13.0cm]{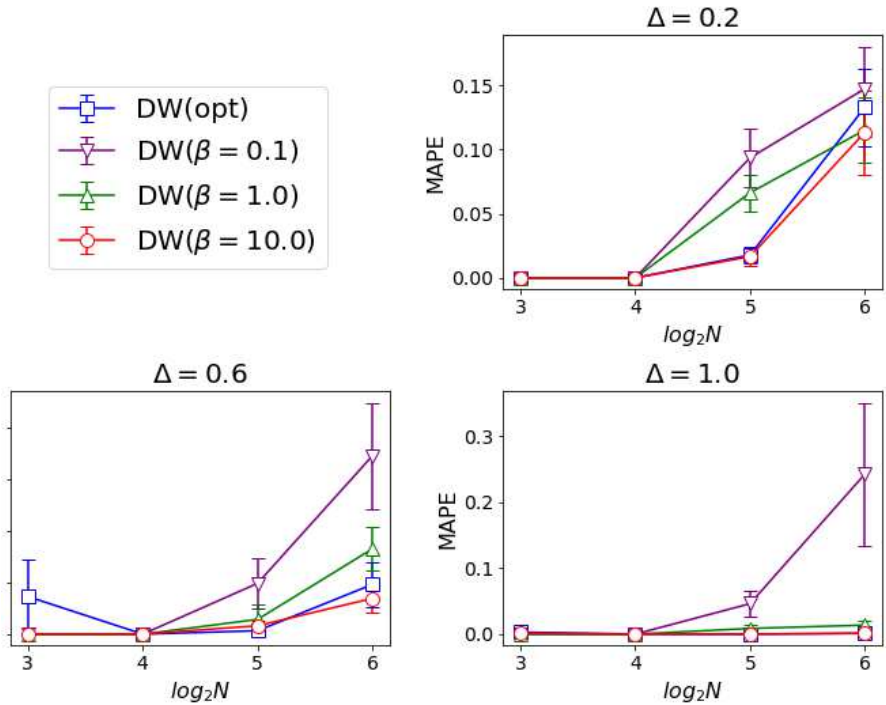}
            \caption{
                $N$-dependence of the MAPEs.
                The squares, lower, upper, and circles correspond to the MAPEs obtained using DW(opt), DW($\beta=0.1$), DW($\beta=1.0$), and DW($\beta=10.0$), respectively.
            }
            \label{fig:QKP_results}
        \end{center}
    \end{figure}
    
    We tested the performance of our algorithm using the quadratic knapsack problem (QKP), which is defined as follows:
    \[
        \begin{aligned}
            \label{eq:KP}
            & \underset{{\bm x}} {\text{maximize}} && \bm{x}^T P \bm{x} \\
            &\text{subject to} && \bm{w}^T \bm{x} \le c
        \end{aligned}
    \]
    where $P = \{ p_{i,j} \}\in \mathbb{Z}_{+}^{N\times N}$ is the profit matrix, $\bm{w} = \{ w_{i} \} \in \mathbb{Z}_{+}^N $ is the weight vector, and $c \in \mathbb{Z} $ is the capacity.
    In this study, we randomly generate $P$ and $\bm{w}$, which was introduced by Gallo $et \ al.$ \cite{QKP_Gallo}.
    The profits $\{ p_{i,j} \}$ are zero with probability $(1-\Delta)$, and non-zero values given by a uniform distribution between 1 and 100 with probability $\Delta$.
    This means that when $\Delta$ is close to 1 (zero), the objective function is given by a random dense (sparse) graph.
    The weights $\{ w_i \}$ are also randomly chosen from $[1, 50]$, and the capacity $c$ is taken from a uniform distribution over $[50, \sum_{i}w_i]$. 
    We generate 10 instances for testing the typical performance of our algorithm.
    To deal with maximization problems on the D-Wave 2000Q, we define the objective function as $f(\bm{x}) = - \bm{x}^T P \bm{x}$.
    
    To study the accuracy, we define the mean absolute percentage error (MAPE) as follows:
    \begin{equation}
        \label{eq:error}
        \text{MAPE} = \frac{1}{N_{\rm inst}}\sum_{k=1}^{N_{\rm inst}} \frac{|f_k(\bm{x}^*_{\rm opt})-f_k(\bm{x}^*_{\rm feas})|}{f_k(\bm{x}^*_{\rm opt}),}
    \end{equation}
    where $f_k(\bm{x})$ is the objective function for the $k$th instance, and $N_{\rm inst}$ is the total number of instances.
    Here, $\bm{x}^*_{\rm opt}$ is the optimal solution obtained using the Gurobi optimizer \cite{Gurobi}, and $\bm{x}^*_{\rm feas}$ is the feasible solution in the ADMM.
    Thus, MAPE = 0 corresponds to the ADMM achieving the optimal solutions for all instances.
    We study the MAPEs obtained using optimization (DW(opt)) and sampling (DW($\beta$)) modes with $\beta = 0.1$, $1.0$, and $10.0$.
    During these experiments, we set the annealing time to $20$ $\mu$s and generate 2000 samples.
    To check the performance of our algorithm, we calculate the exact solutions using the Gurobi optimizer on a 4-core Intel i7 6700K processor with 64 GB of RAM.
    We set the maximum calculation time in the Gurobi optimizer to 1000 s.
    The predetermined parameters in the ADMM are as follows:
    \begin{subequations}
    \begin{align}
        &\rho = 0.1 \label{eq:rho} \\
        &t_{\rm max} = 30 \label{eq:max} \\
        &t_{\rm conv} = 10 \label{eq:conv} \\
        &\epsilon = 10^{-3}. \label{eq:epsilon} 
    \end{align}
    \end{subequations} 
    
    Fig.\ref{fig:QKP_results} shows the $N$-dependence of the MAPEs in $\Delta = 0.2, 0.6, 1.0$.
    As shown in Fig.\ref{fig:QKP_results}, we attain the feasible solutions for all instances at up to $N = 64$.
    The results demonstrate the superiority of our ADMM approach.
    If we use the slack variables, the D-Wave 2000Q cannot solve the problems at $N = 64$ because of the additional binary variables.
    The ADMM allows us to deal with larger-size problems on the D-Wave 2000Q than allowed by the previous approach.

    We compared the MAPEs obtained by DW(opt), DW($\beta=0.1$), DW($\beta=1.0$), and DW($\beta=10.0$).
    Fig.\ref{fig:QKP_results} shows that all MAPEs increase with an increase in $N$ in $\Delta=0.2$.
    By contrast, for $\Delta=1.0$, the MAPEs by DW(opt), DW($\beta=1.0$), and DW($\beta=10.0$) remain at near zero even when $N$ increases. 
    These results indicate that the ADMM can accurately find feasible solutions for the QKP on a dense graph.
    Table \ref{tab:QKP_results_N=64} shows the $\Delta$-dependence of the MAPEs for $N = 64$.
    The MAPEs in DW(opt), DW($\beta=0.1$), DW($\beta=1.0$), and DW($\beta=10.0$) decrease as $\Delta$ increases.
    DW($\beta=10.0$) outperforms the other postprocessing at $N = 64$.
    In addition, the accuracy of DW(opt) is comparable to that of DW($\beta=10.0$).

    \begin{table}[h]
        \center
        \small
        \caption{$\Delta$-dependence of MAPEs at $N = 64$.}
        \begin{tabular}{|Y|Y|Y|Y|Y|} \hline
            & optimization & $\beta=0.1$ & $\beta=1.0$ & $\beta=10.0$ \\ \hline \hline
            $\Delta=0.2$ & 0.1329 & 0.1473 & 0.1152 & 0.1133 \\ \hline
            $\Delta=0.6$ & 0.0192 & 0.0690 & 0.0330 & 0.0138 \\ \hline
            $\Delta=1.0$ &  0.0014 & 0.2423 & 0.0136 & 0.0012 \\ \hline
        \end{tabular}
        \label{tab:QKP_results_N=64}
    \end{table}

\begin{figure}[htbp]
        \begin{center}
            \includegraphics[clip, width=12.0cm]{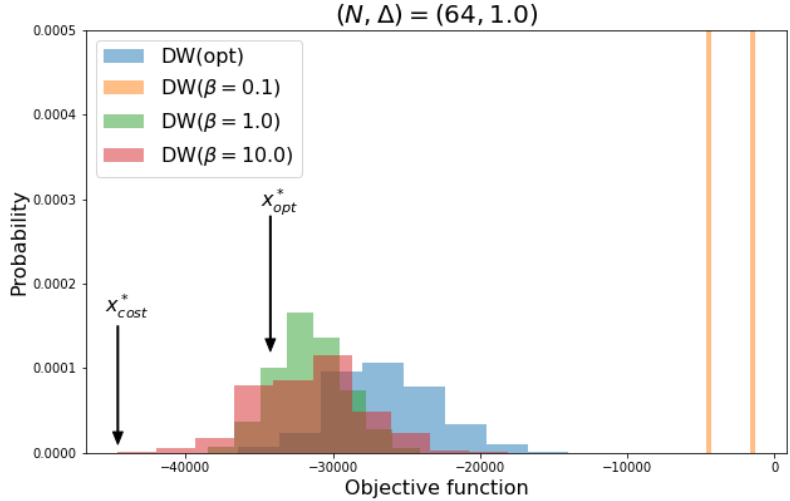}
            \caption{
                Histograms obtained using DW(opt), DW($\beta=0.1$), DW($\beta=1.0$), and DW($\beta=10.0$).
                The horizontal axis is the objective function, and we show the vertical axis with only $[0.0, 0.0005]$.
            }
            \label{fig:dist}
        \end{center}
    \end{figure}

    The difference in accuracy between the postprocessing modes can be explained by the efficient sampling near the lowest-cost solution $\bm{x}_{\rm cost}^*$.
    Fig.\ref{fig:dist} shows the histograms for the instance when the ADMM is finished in $(\Delta,N)=(1.0,64)$.
    Here, the horizontal axis corresponds to the objective function $f(\bm{x})$.
    As can be seen from Fig.\ref{fig:dist}, $\bm{x}_{\rm opt}^*$ is near $\bm{x}_{\rm cost}^*$, and does not correspond to the one.
    Therefore, to find an optimal or accurate solution, sampling near $\bm{x}_{\rm cost}^*$ is necessary.
    In fact, DW(opt), DW($\beta=1.0$), and DW($\beta=10.0$) have broad histograms located near $\bm{x}_{\rm cost}^*$, and are successful in finding $\bm{x}_{\rm opt}^*$. 
    For this reason, DW(opt) and DW($\beta=10.0$) show an accurate performance in our QKP experiments.

    Here, we comment on the $\beta$-dependence of the sampling mode.
    If the samples are indeed generated from the Boltzmann distribution, we obtain the widely spread histogram with $\beta=0.1$.
    However, DW($\beta=0.1$) has spike-like distributions that are far from $\bm{x}_{\rm cost}$.
    The reason for this is not clarified because we cannot access the postprocessing on D-Wave 2000Q.
    Therefore, we should be careful in tuning $\beta$ of the sampling mode.
    From our results, we recommend using DW($\beta = 10.0$) for the QKP.

    \section{\label{sec:discussion}Discussion}
    
        \begin{figure}[htbp]
            \begin{center}
                \includegraphics[clip,width=13.0cm]{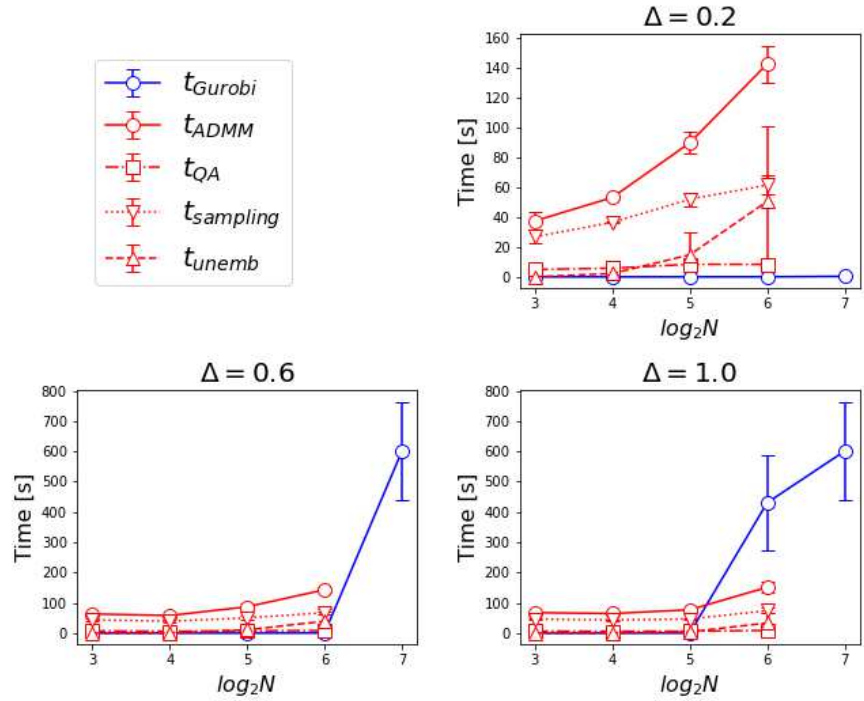}
                \caption{
                    $N$-dependence of the computation times in $\Delta=0.2, 0.6, 1.0$.
                    The red and blue circles represent the total computation times by the ADMM and Gurobi optimizer, respectively.
                    We also show the total QA times with the red squares.
                    The red lower and upper triangles correspond to the total sampling and unembedding times, respectively.
                }
                \label{fig:time}
            \end{center}
        \end{figure}

        We discuss improving the accuracy of the ADMM.
        We obtain $\epsilon_{\rm ave}>0.0$ in our experiments, which means that the ADMM cannot achieve the optimal solutions for several instances.
        A simple way to improve the accuracy is by tuning the value of $t_{\rm conv}$.
        In this study, we terminate the ADMM when the feasible solution is not improved in $t_{\rm conv}$ steps.
        We can obtain more accurate solutions by iterating more ADMM updates.
        Another way is to generate more samples on the D-Wave 2000Q.
        Because D-Wave 2000Q is a stochastic sampler, we need many samples to obtain an optimal solution.

        We compare the computation times obtained by the ADMM and Gurobi optimizer.
        We define the total QA, sampling, unembedding with $t_{\rm QA}$, $t_{\rm sampling}$, and $t_{\rm unemb}$, respectively.
        The $t_{\rm QA}$ value is the total access time for the quantum processing unit on the D-Wave 2000Q.
        The total sampling time $t_{\rm sampling}$ involves the Internet latency, $t_{\rm QA}$, and other processing on the D-Wave 2000Q.
        In addition, $t_{\rm unemb}$ is the total unembedding time in the ADMM steps.
        We set the total computation times by the ADMM and Gurobi optimizer as $t_{\rm ADMM}$ and $t_{\rm Gurobi}$, respectively.
        Here, $t_{\rm ADMM}$ is given as the summation of $t_{\rm QA}$, $t_{\rm sampling}$, $t_{\rm unemb}$, and other processes on the CPU.
        Fig.\ref{fig:time} shows the $N$-dependence of $t_{\rm Gurobi}$ and $t_{\rm ADMM}$.
        The red and blue circles show the instance-averaged computation time, $t_{\rm ADMM}$ and $t_{\rm Gurobi}$, obtained using the DW($\beta=10.0$) and Gurobi optimizer, respectively.
        We conducted a simulation using the Gurobi optimizer at up to $N = 128$.
        Fig.\ref{fig:time} also shows $t_{\rm QA}$, $t_{\rm sampling}$, and $t_{\rm unemb}$ with the red squares and lower and upper triangles, respectively.
        In $\Delta=0.2$ and $0.6$, the Gurobi optimizer is significantly faster than the ADMM.
        However, $t_{\rm Gurobi}$ increases dramatically as $\Delta$ and $N$ increase.
        In fact, we obtain $t_{\rm ADMM} < t_{\rm Gurobi}$ in $(\Delta, N) = (1.0, 64)$.
        Thus, the ADMM can be faster than the exact optimizer with an increase in $\Delta$ and $N$.
        
        Herein, we focus on $t_{\rm QA}$, $t_{\rm sampling}$, and $t_{\rm unemb}$.
        As can be seen from Fig.\ref{fig:time}, $t_{\rm sampling}$ and $t_{\rm unemb}$ grow as $N$ increases, whereas $t_{\rm QA}$ remains almost constant.
        Therefore, $t_{\rm ADMM}$ can be much faster if we reduce the computational overhead, such as $t_{\rm sampling}$ and $t_{\rm unemb}$.
        In particular, the embedding and unembedding techniques are necessary only when the D-Wave 2000Q implements a sparse graph.
        Thus, our method can outperform the exact optimizers if a quantum annealer on a larger and denser graph is developed in the future.

    \section{\label{sec:summary}Summary}
        In this study, we reported a new algorithm for solving inequality-constrained binary optimization using the D-Wave 2000Q.
        We defined the new cost function with the augmented Lagrangian method and developed a hybrid algorithm that combines QA and ADMM.
        We tested the performance of our algorithm for QKP and obtained three significant results.
        First, our algorithm finds feasible solutions for large-sized problems that cannot be computed using a previous approach.
        Next, the denser the coupling of the logical variables is, the more accurately the ADMM can find the feasible solutions.
        Finally, the optimization or sampling mode with $\beta=10.0$ is appropriate for our ADMM algorithm.
        We also compare the computation times obtained by the ADMM and the exact optimizer.
        We show that the ADMM can be faster than the exact optimizer when a QKP is given on a large and dense graph.

\bibliographystyle{unsrt}
\bibliography{reference}
\end{document}